\documentclass{PoS}

\usepackage{cite}
\usepackage{amsmath}

\title{Infrared singularities in multi-leg scattering amplitudes}

\ShortTitle{IR singularities in multi-leg amplitudes}

\author{\speaker{Einan Gardi}\\
        Higgs Centre for Theoretical Physics, School of Physics and Astronomy, \\
The University of Edinburgh,  
Edinburgh EH9 3JZ, Scotland, UK\\
        E-mail: \email{Einan.Gardi@ed.ac.uk}}

\abstract{I discuss the state-of-the-art knowledge of long-distance singularities in multi-leg gauge-theory scattering amplitudes and report on an on-going calculation of the three-loop soft anomalous dimension through the renormalization of correlators of semi-infinite Wilson lines. I  also discuss the non-Abelian exponentiation theorem that has been recently generalised to multiple Wilson lines and demonstrate its application in computing the soft anomalous dimension. Finally, I present recent results for multiple-gluon-exchange webs and discuss their analytic structure.}

\FullConference{ Loops and Legs in Quantum Field Theory - LL 2014,\\
		 27 April - 2 May 2014 \\
		 Weimar, Germany }

\begin{document}

\section{Infrared singularities}

Long-distance singularities are a salient characteristic of gauge-theory scattering amplitudes. I~will report on recent progress in studying these singularities in the context of multi-leg amplitudes. The first part of the talk is a brief review of what we know and what we wish to know about infrared singularities. I will then explain the strategy for computing the three-loop soft anomalous dimension and report on an on-going computation. In the second part of the talk I review the non-Abelian exponentiation theorem, and then explain how it helps in computing the contributions to the anomalous dimension from a special class of diagrams, ones involving only gluon exchanges between the Wilson lines.

Let us begin by summarising what is known about the singularity structure of massless scattering amplitudes. We consider an $L$-leg amplitude depending on the scales $s_{ij}=2p_i\cdot p_j$, and analyse it in the limit where $s_{ij}$ are all large compared to the QCD scale (fixed-angle scattering).  We use dimensional regularization, $D=4-2\epsilon$, with $\epsilon<0$. Long-distance singularities emerge from two distinct regions in the integration over gluon loop momenta: soft (all momenta components are small compared to $\{s_{ij}\}$) and collinear with respect to one of the hard external legs. Both of these factorize, such that 
\begin{align}
\label{eq:factorization}
{\cal M}(p_i)=\exp\left\{-\frac12 \int_0^{\mu^2}\frac{d\lambda^2}{\lambda^2}
\Gamma(\lambda^2/s_{ij},\alpha_s(\lambda^2,\epsilon))\right\}\,{\cal H}(p_i)
\end{align}
where the hard function ${\cal H}(p_i)$ is finite for $\epsilon=0$, while the exponential factor includes all the singularities. These are generated through the integration over the $D$-dimensional coupling $\alpha_s(\lambda^2,\epsilon)$, which extends down to $\lambda^2=0$. The anomalous dimension function $\Gamma(\lambda^2/s_{ij},\alpha_s(\lambda^2,\epsilon))$ is a finite function, which encodes soft and collinear singularities, to all orders in the coupling. A simple ansatz for this quantity -- the so-called \emph{Dipole formula} -- has been proposed a few years ago~\cite{Becher:2009cu,Gardi:2009qi,Becher:2009qa,Dixon:2009gx,Dixon:2009ur}, 
\begin{align}
\label{dipole}
\Gamma_{\text{Dip.}}(\lambda^2/s_{ij},\alpha_s)=
 \, \frac14 \,
 \widehat{\gamma}_K\left(\alpha_s  \right) 
\sum_{(i,j)} \, \ln\left(\frac{\lambda^2}{-s_{ij}}\right) \,  \mathrm{T}_i \cdot   \mathrm{T}_j \, 
   + \, \sum_{i=1}^n \, \gamma_{J_i} \left(\alpha_s \right)\,.
\end{align}
Hard-collinear singularities are generated by the anomalous dimension $\gamma_{J_i}$, which depends on the spin of the hard parton $i$, but is however, colour singlet, and thus does not present any increasing complexity in the multi-leg case as compared to the simplest amplitudes. These collinear singularities can also be extracted from the massless form factor~\cite{Dixon:2008gr}, which is known to three loops~(e.g.~\cite{Moch:2005tm,Gehrmann:2010ue}). In contrast, soft singularities may a priori present complicated correlations between colour and kinematic degrees of freedom in the entire process.  Yet, according to the dipole formula (\ref{dipole}) soft singularities are remarkably simple: they are effectively associated with two-body interactions (colour dipoles) between any pair of hard partons, having momenta $p_i$ and $p_j$, and colour charges ${\rm T}_i$ and ${\rm T}_j$, respectively. 

This dipole ansatz is based on an explicit calculation~\cite{Aybat:2006wq,Aybat:2006mz} which completely fixed the singularity structure at two loops, consistently with previous predictions~\cite{Catani:1998bh,Sterman:2002qn}. The all-order generalization (\ref{dipole}) was inspired by constraints based on soft-collinear factorization and the invariance of soft singularities with regards to momentum rescaling~\cite{Becher:2009cu,Gardi:2009qi,Becher:2009qa}. These constraints can be formulated as a set of first-order linear differential equations for the soft anomalous dimension, where the inhomogeneous term is fixed by the lightlike cusp anomalous dimension $\gamma_K^{(i)}\left(\alpha_s  \right)$. Eq.~(\ref{dipole}) is a particular solution to these constraints, provided that $\gamma_K^{(i)}\left(\alpha_s  \right)$ admits Casimir scaling, namely it is proportional to the quadratic Casimir in the relevant representation, $\gamma_K^{(i)}\left(\alpha_s  \right)={\rm T}_i\cdot{\rm T}_i\,\,\widehat{\gamma}_K\left(\alpha_s  \right)$, a property that may not hold beyond three loops. 
It follows that potential corrections going beyond (\ref{dipole}) fall into two categories: (1)  corrections associated with quartic (or higher) Casimir operators, which contribute to the cusp anomalous dimension and thus modify the non-homogeneous constraint equations at four or higher loop order; and (2) corrections that depend on conformally invariant cross ratios, $\rho_{ijkl}=\frac{(p_i\cdot p_j)\, (p_k\cdot p_l)}{(p_i\cdot p_k)\, (p_j\cdot p_l)}$, and therefore satisfy the homogeneous set of equations, and then 
$\Gamma=\Gamma_{\rm Dip.}+\Delta(\rho_{ijkl})$ with $\Delta(\rho_{ijkl})={\cal O}(\alpha_s^3)$.
Only the latter correction may appear at the three-loop order. Determining
the correction function $\Delta(\rho_{ijkl})$ is the main motivation for the calculation I will be discussing in the next section.

Much work has been done in recent years on further constraining $\Delta(\rho_{ijkl})$~\cite{Dixon:2008gr,Gardi:2009qi,Dixon:2009gx,Becher:2009cu,Becher:2009qa,Dixon:2009ur,Gardi:2009zv,Gehrmann:2010ue,Bret:2011xm,DelDuca:2011ae,Ahrens:2012qz,Naculich:2013xa,Caron-Huot:2013fea}. Indeed, as a consequence of the non-Abelian exponentiation theorem, this function must have a colour factor of the form $f^{abe}f^{cde}{\rm T}_1^a{\rm T}_2^b{\rm T}_3^c{\rm T}_4^d$ (and permutations). This, in combination with Bose symmetry, implies a strong constraint on the kinematic dependence. Furthermore, it has been shown that $\Delta(\rho_{ijkl})$ is further constrained by the known behaviour of multi-leg amplitudes in collinear limits and in the Regge limit. Nevertheless, it appears that general considerations fall short of determining $\Delta(\rho_{ijkl})$, and an explicit three-loop calculation is required. 

It is useful to contrast the singularity structure of massless scattering amplitudes with that of massive ones. The latter is important in the context of heavy quark production, but it is also interesting from a theoretical perspective. Clearly for heavy partons there are no collinear singularities; interestingly, the pattern of soft singularities is also radically different: here the soft anomalous dimension does not take a dipole form, but instead, already at two loops, it acquires a tripole component with a colour factor of the form $f^{abc}{\rm T}_1^a{\rm T}_2^b{\rm T}_3^c$~\cite{Mitov:2009sv,Becher:2009kw,Beneke:2009rj,Czakon:2009zw,Ferroglia:2009ep,Ferroglia:2009ii,Chiu:2009mg,Mitov:2010xw,Ferroglia:2010mi,Chien:2011wz}.
\begin{figure}[htb]
\begin{center}
\scalebox{.7}{\includegraphics{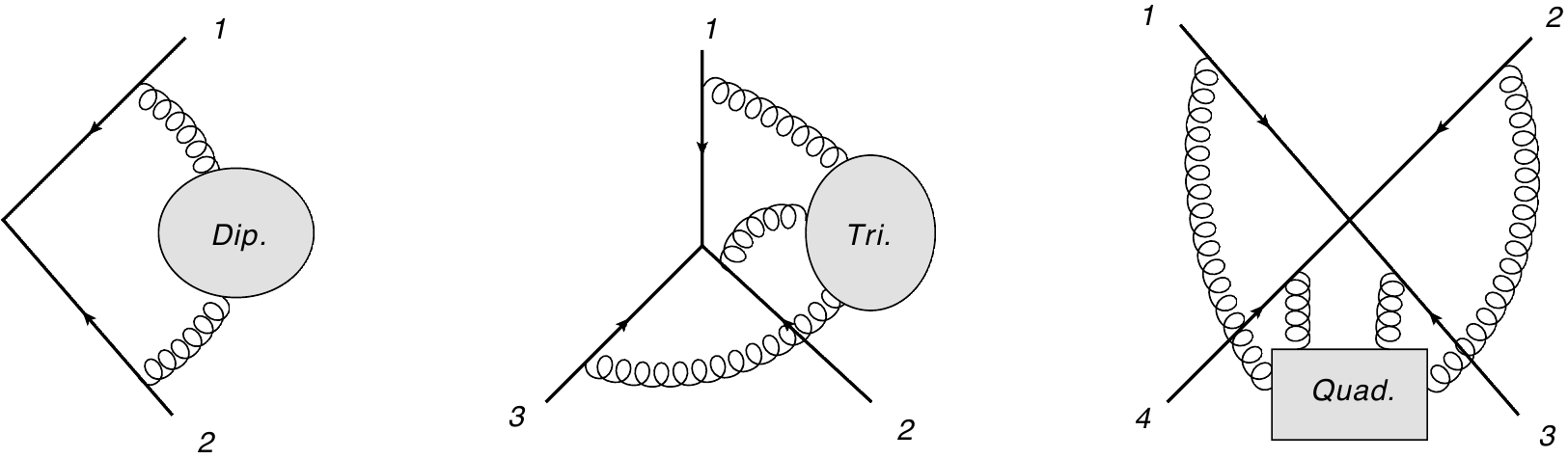}}
\caption{Dipole, tripole and quadrupole interactions involving two, three and four hard partons respectively. The dipole interaction (left) -- ${\rm T}_1\cdot {\rm T}_2$ -- governs soft singularities in massless amplitudes, at least through two loops. The tripole form (middle) -- $f^{abc}{\rm T}_1^a{\rm T}_2^b{\rm T}_3^c$, which is forbidden in the massless case to all loops, appears in the massive soft anomalous dimension already at two loops. The quadrupole interaction (right) -- $f^{abe}f^{cde}{\rm T}_1^a{\rm T}_2^b{\rm T}_3^c{\rm T}_4^d$ (and permutations) -- is the form of the potential three-loop correction to the soft anomalous dimension in the massless case.}
\label{Connected_four}
\end{center}
\end{figure}

\section{The soft anomalous dimension from correlators of Wilson lines\label{sec:correlator}}

Let us now turn to describe our general strategy for computing soft singularities. One possibility is to compute an on-shell multi-leg scattering amplitude in dimensional regularization and examine directly its infrared singularities. This is difficult to do at three loops, but it is also clear that the level of complexity of the finite parts of an amplitude is far greater than that of its singularities. Our aim is to determine singularities directly and elucidate their simple, universal structure. This can be done by considering the eikonal approximation, or equivalently by replacing all hard partons by  semi-infinite Wilson lines, and computing correlators of the form:
\begin{equation}
\label{S}
{\cal S}=\exp\left[ w \right]= \left<\Phi_{\beta_1}\,\otimes\,\Phi_{\beta_2}\,\otimes\,\ldots \Phi_{\beta_L}\right>\,;\qquad\quad
\Phi^{(l)}_{\beta_l}\,\equiv\,
{\cal P}\exp\left[{\mathrm i}g_s \int_0^{\infty}dt \beta_l\cdot {A}(t\beta_l)\right]\,,
\end{equation}
where each Wilson line $\Phi_{\beta_i}$ with 4-velocity $\beta_i^{\mu}$ represents a hard parton with momentum $p_i^{\mu}$, where $\beta_i^{\mu}$ is proportional to $p_i^{\mu}$. 
Soft interactions correspond to emission from the Wilson lines (hard collinear singularities are not captured by this approximation, and need to be accounted for separately), while the hard interaction is replaced by a local vertex where the Wilson lines all meet, and where an arbitrary colour exchange may take place. This vertex gives rise to new ultraviolet singularities (that are unrelated to the renormalization of the theory). Importantly, for any number of non-lightlike Wilson lines, this vertex is multiplicatively renormalizable~\cite{Polyakov:1980ca,Arefeva:1980zd,Dotsenko:1979wb,Brandt:1981kf}.
Working in pure dimensional regularization, radiative corrections to correlators of semi-infinite Wilson lines (\ref{S}) vanish: they all result in scale-less integrals. As a consequence the renormalized correlator is fully determined by its renormalization factor, ${\cal S}_{\rm ren.}=Z$. Thus, the problem of computing infrared singularities has been mapped to the problem of renormalizing the Wilson-line correlator, hence the notion of 
the soft anomalous dimension, $\Gamma_{\cal S}=-Z^{-1}dZ/d\ln\mu^2$.
This mapping was noted already in \cite{Korchemsky:1985xj} in the context of the cusp anomalous dimension, where the correlator consists two Wilson lines; it has since been extensively used for both the two-line and the multi-line cases. It is also the framework we use in computing the three-loop anomalous dimension as described below. 

Importantly, we take the Wilson lines all non-lightlike, $\beta_i^2\neq 0$. This has several advantages: firstly it avoids collinear singularities, restricting to one divergence per order, secondly this guarantees multiplicative renormalizability. Finally, by performing the general non-lightlike Wilson line calculation we gain information about infrared singularities of amplitudes with heavy partons. The lightlike limit is a special case of this result; the limit may be taken at the end. Note that upon working with non-lightlike Wilson lines, the symmetry of the Feynman rules with respect to rescaling any of the velocities $\beta_i^{\mu}$, along with Lorentz invariance, implies that the kinematic dependence will be on the normalised scalar products between the 4-velocities, $\gamma_{ij}=2\beta_i\cdot \beta_j/\sqrt{\beta_i^2\beta_j^2}$.

\section{Computing the three-loop soft anomalous dimension~\label{sec:connected}}

\begin{figure}[htb]
\begin{center}
\scalebox{.5}{\includegraphics{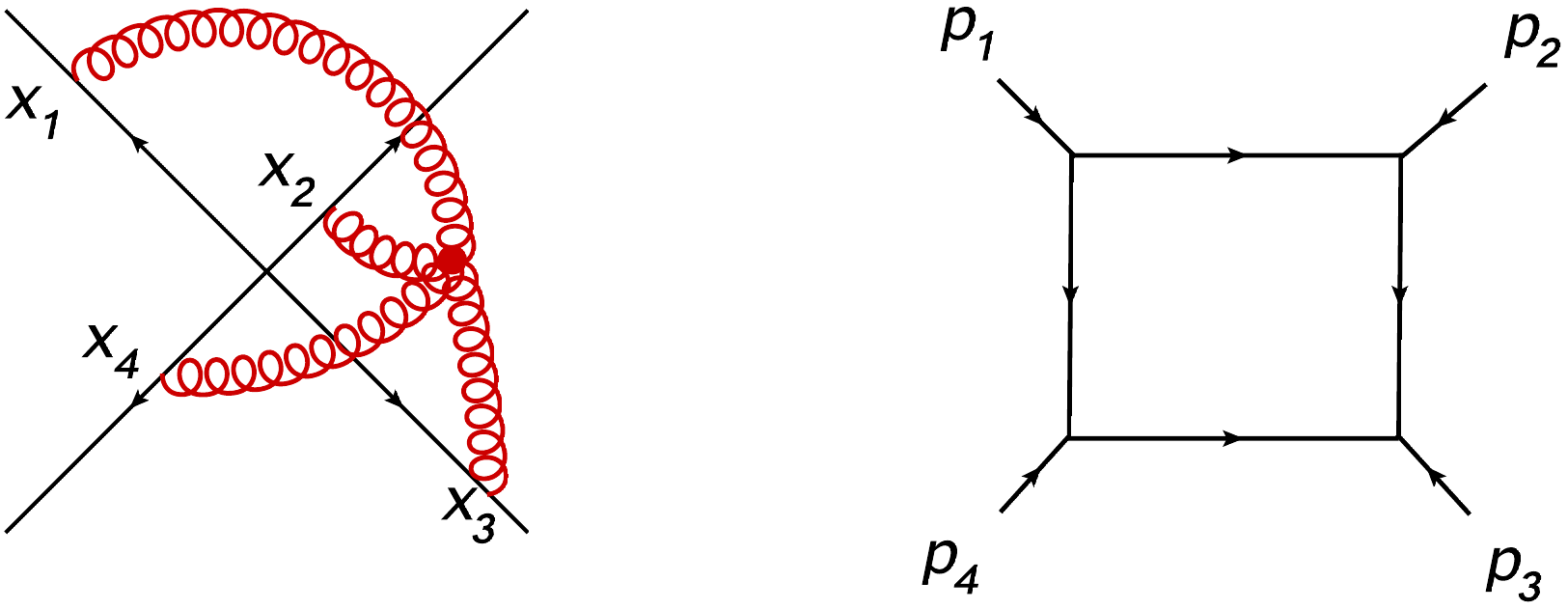}}\\
\vspace*{20pt}
\scalebox{.5}{\includegraphics{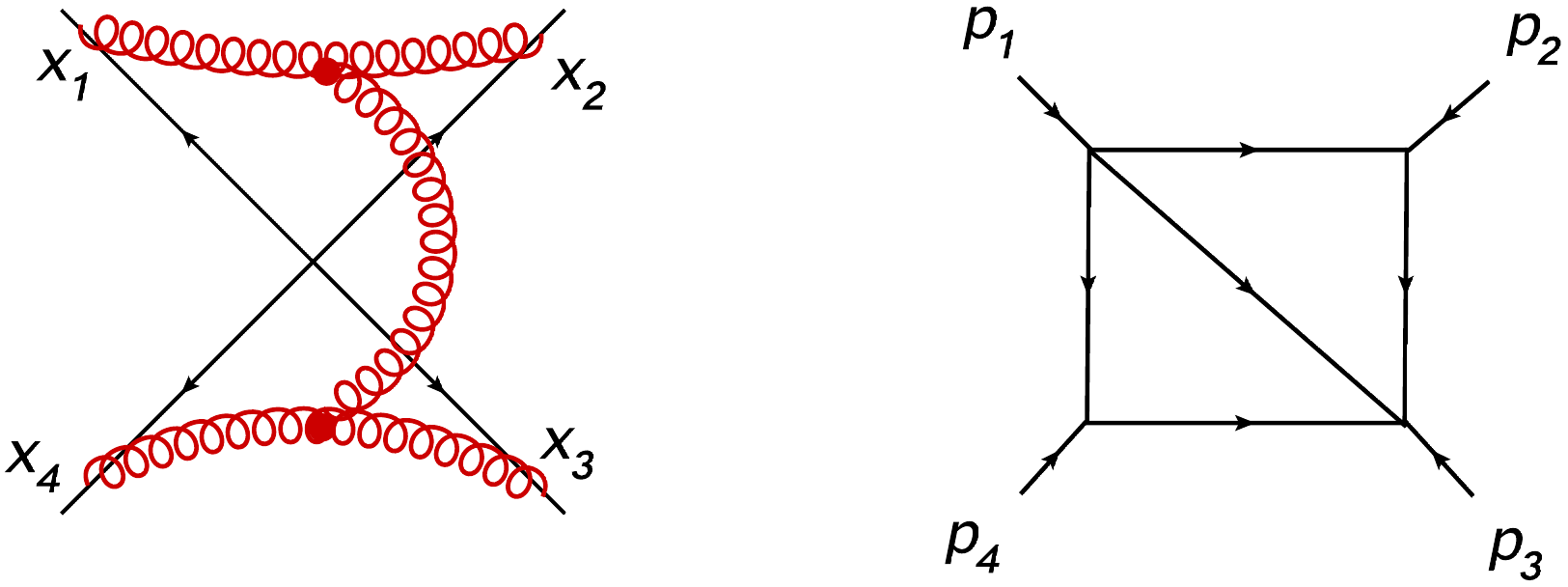}}
\caption{Connected graphs contributing to the three-loop soft anomalous dimension with a quadrupole colour factor $f^{abe}f^{cde}{\rm T}_1^a{\rm T}_2^b{\rm T}_3^c{\rm T}_4^d$ (and permutations), and their mapping to an auxiliary momentum space. Left: the two types of connected graphs involving four semi-infinite Wilson lines with 4-velocities $\beta_1$ through $\beta_4$; a single gluon is emitted from each of the Wilson lines at positions $x_1$ through $x_4$, respectively. The four gluons are connected to the Wilson lines via a single 4-gluon vertex (top) or two 3-gluon vertices (bottom). Right: a four-mass one-loop box  integral (top) and a four-mass two-loop diagonal box integral (bottom) corresponding to the integration over the position of the 4 and 3 gluon vertices in the diagrams on the left.}
\label{4g_connected}
\end{center}
\end{figure}

\begin{figure}[htb]
\begin{center}
\scalebox{.6}{\includegraphics{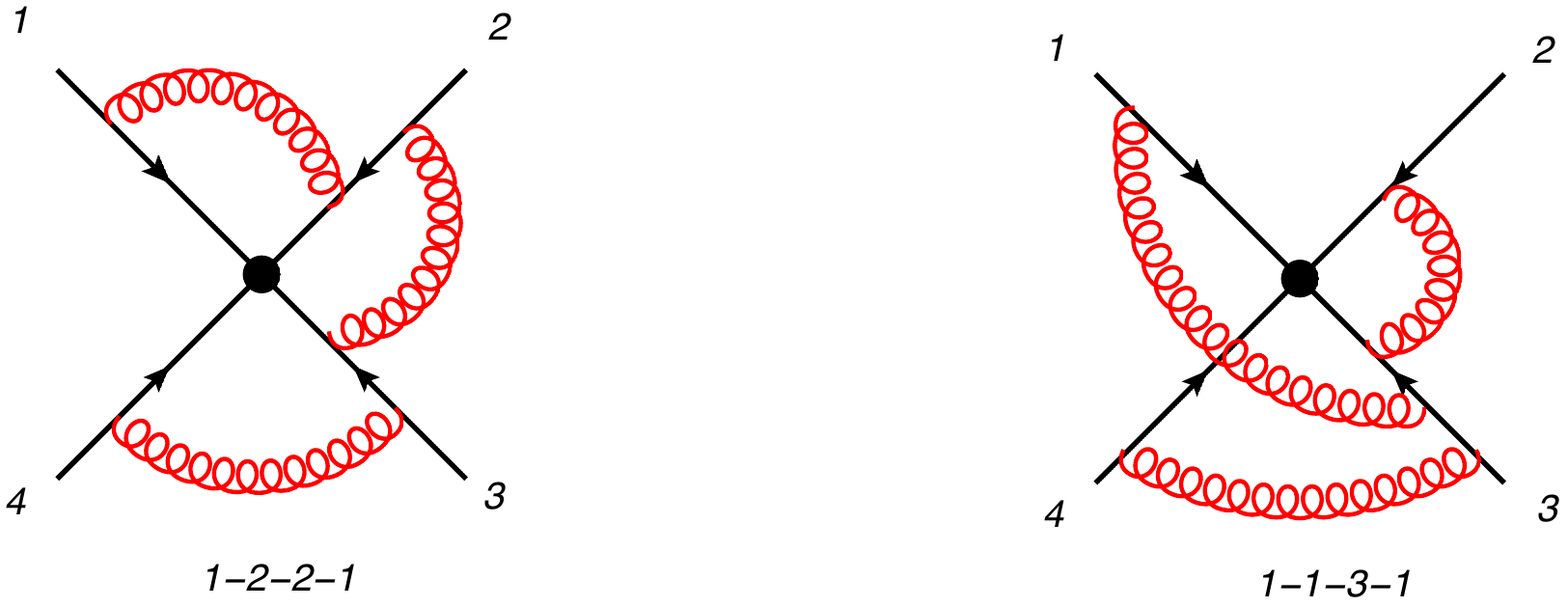}}
\scalebox{1.}{\includegraphics{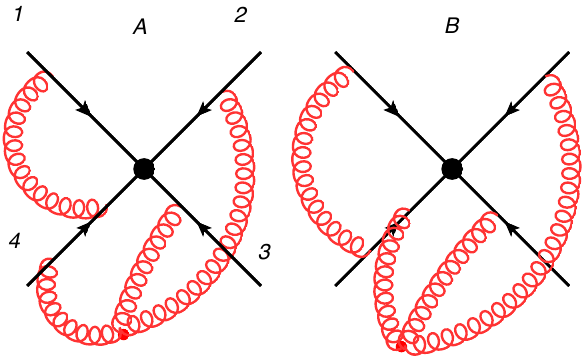}}
\caption{Three webs contributing to the soft anomalous dimension connecting four lines at three-loops with the quadrupole colour factor of the type $f^{abe}f^{cde}{\rm T}_1^a{\rm T}_2^b{\rm T}_3^c{\rm T}_4^d$. 
Top left: one of the four diagrams of the multi-gluon-exchange 1-2-2-1 web (the other three diagrams are related to the one shown by permuting the gluon attachments to lines 3 and 4, see below). Top Right: one of six diagrams of the multi-gluon-exchange 1-1-3-1 web. Bottom: the two diagrams of the 1-1-1-2 web.}
\label{mge_and_1112}
\end{center}
\end{figure}

As discussed above, soft singularities are known at the two-loop level for amplitudes with any number of either massless or massive legs. Here I report on an on-going calculation aimed at determining the soft anomalous dimension $\Gamma_{\cal S}$ at three loops. As a first priority we focus on four-parton correlations in the massless case, the function $\Delta(\rho_{ijkl})$ mentioned above; we consider semi-infinite non-lightlike Wilson lines (i.e. we take $\beta_i^2\neq 0$, expanding in $1/\gamma_{ij}$ at the end) and perform the entire calculation in configuration space, using dimensional regularization with $D=4-2\epsilon$ and $\epsilon>0$. The relevant graphs 
are shown in figures \ref{4g_connected} and \ref{mge_and_1112}.  All these diagrams, or webs\footnote{Webs, generally, are diagrams that contribute to the exponent; the terminology will be explained more precisely in section \ref{sec:non-AbelianExp} below.} contribute to $\Gamma_{\cal S}$ with a colour factor of the form $f^{abe}f^{cde}{\rm T}_1^a{\rm T}_2^b{\rm T}_3^c{\rm T}_4^d$ (and permutations). 
We note that while all these diagrams are required to get the complete, gauge-invariant result, working in the Feynman gauge, it is natural to make a further separation of the diagrams into three classes based on the number of connected subdiagrams\footnote{In this context, as in the formulation of non-Abelian exponentiation below, the connectedness property refers to the graph without the Wilson lines.}, which directly dictates the analytic structure\footnote{Specifically, it dictates the structure of the rational function multiplying the polylogs. For supersymmetric Wilson lines in ${\cal N}=4$, the same classification is associated with polynomial dependence on the cosine of the angles between the orientations of the scalar field in the internal space associated with the different Wilson lines ~\cite{Correa:2012nk,Henn:2012qz,Henn:2013wfa}.} 
of the result~\cite{Gardi:2013saa}.  
The results for the 1-2-2-1 and 1-1-3-1 webs in figure~\ref{mge_and_1112}, each having three individual gluon exchanges, will be briefly discussed in section \ref{sec:mge} below; these results
have been obtained last year~\cite{Gardi:2013saa}. 
The calculation of the \hbox{1-1-1-2} web, which has two connected subdiagrams, one containing a three-gluon vertex, requires an extended set of techniques; the result will be published soon~\cite{Falcioni:2014tba}. 
In parallel the computation of the lightlike limit of connected webs (figure~\ref{4g_connected}) which uses entirely different techniques, is nearing completion~\cite{Almelid:2013tb}. In the following we briefly discuss this calculation. 
 
Before describing the computation of the rather challenging integrals of figure~\ref{4g_connected}, it is useful to review the calculation of the simplest radiative correction to a product of semi-infinite Wilson lines, namely that of a single gluon exchange\footnote{Since a single gluon only connects two Wilson lines, this is essentially the one-loop correction to the angle-dependent cusp anomalous dimension. We note in passing that the latter has been computed to two loops already in 1987~\cite{Korchemsky:1987wg} (see also~\cite{Kidonakis:2009ev}). Recently there has been significant progress towards higher loops~\cite{Correa:2012nk,Henn:2012qz,Henn:2013wfa} including the complete three-loop and a partial four-loop result in ${\cal N}=4$ Super Yang-Mills. Furthermore, in this conference Henn reported on new results at three loop in QCD~\cite{Grozin:2014axa}.}.
The calculation is particularly simple in configuration space: 
we parametrise the point of emission along Wilson line 1 as $x_1^{\mu}=\beta_1^{\mu} s$ and along Wilson line 2 as $x_2^{\mu}=\beta_2^{\mu} t$; the gluon propagator in $D=4-2\epsilon$ space-time dimensions is $D_{\mu\nu}(x_1-x_2)=-{\cal N} g_{\mu\nu} (-(x_1-x_2)^2+{\rm i}0)^{\epsilon-1}$ with ${\cal N}=\Gamma(1-\epsilon)/(4\pi^{2-\epsilon})$, so the diagram evaluates to:
\begin{align}
\label{one-loop}
\begin{split}
W^{(1)}(\gamma_{ij},\epsilon)
  &= T_i\cdot T_j \, g_s^2\,{\cal N}\, \beta_i\cdot\beta_j\!
\int_0^{``\infty"}\!\! ds\! \int_0^{``\infty"} \!\! dt
  \, \Big(-(s\beta_i-t\beta_j)^2+{\rm i}0\Big)^{\epsilon-1}
\\
&=\kappa\,T_i\cdot T_j \,\gamma_{ij}\,
\int_0^{\infty} d\sigma \int_0^{\infty}  d\tau
  \Big(\sigma^2+\tau^2-\gamma_{ij} \sigma \tau-{\rm i}0\Big)^{\epsilon-1}
\,{\rm e}^{-\sigma-\tau}
\\
 &=
\kappa\, T_i\cdot T_j 
 \,\Gamma(2\epsilon) \,\gamma_{ij}\int_0^1dx \, P_{\epsilon}(x,\gamma_{ij})
\end{split}
\end{align}
with $P_{\epsilon}(x,\gamma_{ij})\,\equiv\,\Big[x^2+(1-x)^2-x(1-x)\gamma_{ij}-{\rm i}0\Big]^{\epsilon-1}$ and $\kappa=-(\mu^2/m_{\rm IR}^2)^\epsilon \frac{g_s^2}{2}\,{\cal N}$. A crucial step in setting up this calculation has been to introduce an infrared regulator (implied by the notation ``$\infty$'' in the first line of (\ref{one-loop})). This regulator takes the form of an exponential cut-off along the Wilson lines controlled by $m_{\rm IR}$. In the absence of a cutoff, the result is zero as a scale-less integral. Having introduced the cutoff, dimensional regularization is employed in the ultraviolet, with $\epsilon >0$. The form of the cutoff facilitates straightforward integration with the variables $\lambda=\sigma+\tau$ and $x=\sigma/(\sigma+\lambda)$, where the latter varies between 0 and 1 depending on the angle of emission. The ultraviolet divergence corresponding to $\lambda\to 0$ appears in the last line as a single pole in $\epsilon$, where the coefficient of that pole is independent of the infrared regulator. 

Let us turn now to the integrals of figure~\ref{4g_connected}. Being connected, these graphs have a single ultraviolet pole, similarly to the one-loop correction just discussed. Parametrising the position along each of the Wilson lines by $x_i^{\mu}=\beta_i^{\mu}s_i$ for $i=1..4$ we find that the diagram with a four gluon vertex evaluates to 
\begin{align}
W_{4g}=&g_s^6  \,{\cal N}^4 \, C_{4g} \, \int_0^{``\infty"} ds_1 ds_2 ds_3 ds_4 \,\,{\rm Box}(x_1-x_4,x_2-x_1,x_3-x_2,x_4-x_3) 
\\
C_{4g} \equiv& T_1^a T_2^b T_3^c T_4^d \Big[f^{abe} f^{cde} (\gamma_{13} \gamma_{24}- \gamma_{14}\gamma_{23})+f^{ade}f^{bce}  (\gamma_{12} \gamma_{34}- \gamma_{13}\gamma_{24}) +f^{ace}f^{bde}  (\gamma_{12}\gamma_{34}-\gamma_{14} \gamma_{23}) \Big],\nonumber
\end{align}
where we represented the $D=4-2\epsilon$ dimensional integration over the position of the 4-gluon vertex by ${\rm Box}$. This integral, we claim, is a 4-mass box integral in an auxiliary momentum space, where $p_i=x_i-x_{i-1}$ (see figure~\ref{4g_connected}). Thus the virtualities of the external legs in the box integral are determined by the distances between the positions of the gluon attachments to the Wilson lines. At the next step we wish to perform 
the integral over the overall distance $\lambda=\sum s_i$, which is cut off at infinity by our exponential regulator. This can be easily done, e.g. using the following parametrization: 
{
\begin{align}
\left(\begin{array}{c}
s_1\\
s_2\\
s_3\\
s_4
\end{array}\right)=\lambda \left(\begin{array}{c}
ca\\
c(1-a)\\
(1-c)b\\
(1-c)(1-b)
\end{array}\right)\,,
\end{align}}
given that the Box integral is homogeneous in $\lambda$. At this point the ultraviolet singularity emerges as an overall single pole in $\epsilon$, and $\epsilon$ may then be sent to zero in the Box integral, greatly simplifying the remaining integrations. The second graph in figure~\ref{4g_connected}, involving two 3-gluon vertices, is evaluated in a similar manner, with the main difference being that now, mapping to the auxiliary momentum space yields a four-mass two-loop diagonal box integral, as shown in figure~\ref{4g_connected}. The remaining integrations are non-trivial, and therefore as a first step we performed them to leading order in the expansion near the lightlike limit. This is sufficient for determining the anomalous dimension in the massless case, which depends on  conformally invariant cross ratios of the form
$ 
\rho_{ijkl}=\frac{\gamma_{ij}\,\gamma_{kl}}{\gamma_{ik}\,\gamma_{jl}}
=\frac{(\beta_i\cdot\beta_j)\,(\beta_k\cdot\beta_l)}{(\beta_i\cdot\beta_k) (\beta_j\cdot\beta_l)}
$.
The result~\cite{Almelid:2013tb}, valid up to corrections suppressed by powers of $1/\gamma_{ij}^2$, involves pure functions of weight 5 composed of Goncharov's polylogarithms~\cite{Goncharov.A.B.:2009tja,
Goncharov:2010jf,Duhr:2011zq,Duhr:2012fh}
 with the symbol alphabet ${\cal A}=\big\{ z,\,\bar{z},\,1-z,\, 1-\bar{z},\, 1-z\bar{z},\, z+\bar{z}-z\bar{z},\, 1-z-\bar{z},\,z-\bar{z}\big\}$, where $\rho_{1234}=z\bar{z}$ and
$\rho_{1432}=(1-z)(1-\bar{z})$.

Next we would like to consider other integrals, such as the multi-gluon-exchange webs where three gluons span four Wilson lines. 
As already mentioned these configurations contribute with the same colour factor as the diagrams we just discussed, thus they are part of the same gauge-invariant  sector. A new difficulty arises here due to the fact that these diagrams have subdivergences. 
For example in the top left diagram in figure~\ref{mge_and_1112} (the 1-2-2-1 web), one may consider the limit where the upper gluon shrinks towards the multi-Wilson-line vertex without affecting the other two gluons, then the one on the right and then the bottom one, altogether generating a triple pole in $\epsilon$. 
The same is true for the 1-1-3-1 web, top right in figure~\ref{mge_and_1112}. 
The presence of multiple poles, in turns, implies that the procedure applied so far, namely to determine the coefficient of the $1/\epsilon$ pole would not work well: it would yield a result that depends on the infrared regulator. 
Of course the expectation is that this dependence will cancel out when one sums diagrams, but the result for any individual diagram may be vastly complicated by cut-off artefacts.   
I will show that this problem is entirely avoided by computing together a set of diagrams called a (subtracted) web. 
The underlying reason for this simplification is the workings of non-Abelian exponentiation, which we briefly review in the next section.

\section{The non-Abelian exponentiation theorem\label{sec:non-AbelianExp}}

A central observation, already alluded to in this talk, is that 
soft gluons exponentiate. The exponent $w$ in eq.~(\ref{S}) is simpler than the correlator itself, and it is therefore advantageous to compute it directly.
There are two ways to understand soft gluon exponentiation. The first begins with the observation that the $Z$ factor renormalizing the correlator of Wilson lines in eq.~(\ref{S}) obeys a renormalization-group equation. Upon solving this equation, $Z$ is expressed as an (ordered) exponential of the integral over the anomalous dimension $\Gamma_{\cal S}$, as in eq.~(\ref{eq:factorization}):
\begin{align}
\label{Z}
\begin{split}
Z=& {\bf \rm P}\,\exp\left\{\frac12 \int_{\mu^2}^{\infty}\frac{d\lambda^2}{\lambda^2} \,\Gamma_{\cal S}(\alpha_s(\lambda^2))\right\}
=\,\exp\left\{
\,\frac{1}{2\epsilon}\,\Gamma_{\cal S}^{(1)}\,\alpha_s
+\,\left(\frac{1}{4\epsilon}\,\Gamma_{\cal S}^{(2)}-\frac{b_0}{4\epsilon^2}\,\Gamma_{\cal S}^{(1)}\right)\,\alpha_s^2
\right.\\&
\left.
+\,\left(\frac{1}{6\epsilon}\,\Gamma_{\cal S}^{(3)}
+\frac{1}{48\epsilon^2}\left[\Gamma_{\cal S}^{(1)},\Gamma_{\cal S}^{(2)}\right]-\frac{1}{6\epsilon^2}
\left(b_0\Gamma_{\cal S}^{(2)}+b_1\Gamma_{\cal S}^{(1)}\right)+\frac{b_0^2}{6\epsilon^3}\Gamma_{\cal S}^{(1)}\right)\,\alpha_s^3
\,+\,{\cal O}(\alpha_s^4)\,
 \right\}\,.
\end{split}
\end{align}
Note the higher order poles in the exponent are generated by the running of the coupling. In contrast $Z$ itself has all the way up to $n$-th order poles at order $\alpha_s^n$, even in a conformal theory.
From this perspective it is clear that the structure of the exponent is simpler.
The second, and complementary way to understand exponentiation is the diagrammatic approach: one may consider the diagrammatic rules for computing the exponent $w$ in eq.~(\ref{S}). This point of view, which we shall now briefly summarise (for a more complete review, see~\cite{Gardi:2013jia}), was developed in the context of multiple Wilson lines in recent papers~\cite{Gardi:2010rn,Mitov:2010rp,Gardi:2011wa,Gardi:2011yz,Dukes:2013wa,Dukes:2013gea,Gardi:2013ita,Vladimirov:2014wga}.

Diagrammatic exponentiation was conceived in the context of the Abelian theory in 1961~\cite{Yennie:1961ad}; in this case the exponent $w$ only receives contributions from connected graphs (throughout our discussion `connected' should be understood as referring to the graph after removing the Wilson lines). All non-connected diagrams are reproduced upon expanding the exponential. The next step was taken in the 1980's~\cite{Sterman:1981jc,Gatheral:1983cz,Frenkel:1984pz}, when the non-Abelian exponentiation theorem was first formulated. This was done in the context of a Wilson loop, or two Wilson lines, corresponding to a colour-singlet form factor (for a review see~\cite{Berger:2003zh}).  
The generalization to a product of more than two Wilson lines, as relevant for QCD hard scattering, was only made over the last three years~\cite{Gardi:2010rn,Mitov:2010rp,Gardi:2011wa,Gardi:2011yz,Dukes:2013wa,Dukes:2013gea,Gardi:2013ita}.  

The non-Abelian exponentiation theorem in the two-line case can be phrased as follows: webs, the diagrams that contribute to the exponent, are \emph{irreducible diagrams}. The contribution of any such diagram to the exponent is associated with a modified colour factor, which is the \emph{connected part} of the ordinary colour factor of that diagram~\cite{Frenkel:1984pz,Berger:2003zh}. 
Irreducible graphs are a larger class of graphs as compared to connected graphs: it also includes any non-connected graph whose colour factor cannot be expressed as a product of the colour factors of its subgraphs. Consider for example the diagram involving two crossed gluons; this diagram is irreducible, thus a web. In contrast, the ladder graph, where the two gluons do not cross, is reducible (the colour factor can be expressed as a product of ${\rm T}_1\cdot {\rm T}_2$ for each of the gluons) and is therefore not a web.
Reducible graphs also have subdivergences related to the renormalization of the cusp, while irreducible ones have a single ultraviolet divergence -- this is in direct correspondence with what one expects based on the renormalization properties: at any order in the exponent (\ref{Z}) there is a single $1/\epsilon$ associated with the cusp (here we have excluded running coupling corrections, and in the two-line colour singlet case the commutators vanish). We thus see that each web has a single pole, and the coefficient of that pole can be readily identified as a contribution to the anomalous dimension $\Gamma_{\cal S}^{(n)}$, while reducible diagrams, which have subdivergences, do not contribute.

The generalization of this picture to the case of multiple Wilson lines is non trivial because here many reducible diagrams do in fact contribute to the exponent. It was found~\cite{Gardi:2010rn} that the natural generalization of the concept of webs to the multi-line case is in terms of sets of diagrams: a web is defined as the set of all diagrams which are related to each other by interchanging the order of gluon attachments to (any of) the Wilson lines. 
A given diagram $D$ contributes to the exponent $w$ with an exponentiated colour factor ${\color[rgb]{0.0,0.0,0.0}\widetilde{C}_{D}}$ which is itself a linear combination of the ordinary colour factors of diagrams in the set. The resulting structure is then:
\begin{align}
\label{Mixing}
\begin{split}
{\cal S}=\exp\left[\sum_i W_i\right]\,,
\qquad\qquad
W_i=\sum_{\left\{D\right\}_i}{\cal F}_{D}\, {\color[rgb]{0.0,0.0,0.0}\widetilde{C}_{D}}
=\,\sum_{\left\{D\right\}_i}{\cal F}_{D}\,
\sum_{\left\{D'\right\}_i}\,{\color[rgb]{0.0,0.0,0.0}R_{DD'}}\,\,C_{D'}\,= {\cal F}^T {\color[rgb]{0.0,0.0,0.0}R} C\,
\end{split}
\end{align}
where ${\cal F}_{D}$ and $\,C_{D}$ are respectively the kinematic integral and ordinary colour factor associated with diagram $D$, and 
$R$ is a matrix whose entries are rational numbers. 
We refer to $R$ as the \emph{web mixing matrix}. 
Using a path-integral formulation~\cite{Laenen:2008gt} and the replica trick from statistical physics, an algorithm was formulated~\cite{Gardi:2010rn} allowing one to compute $R$ for any given web. The mixing matrix has a rich combinatorial structure~\cite{Gardi:2011wa,Gardi:2011yz,Dukes:2013wa,Dukes:2013gea} and some interesting properties relating to the physics of soft gluons. An important property is that $R$ is idempotent, $R^2=R$, for any web. This means that it is always diagonalisable, and has eigenvalues that are either 0 or 1. Its action can be understood as a projection operator acting in the vector space of the colour factors of the diagrams in the web: it selects particular linear combinations of these colour factors, the ones corresponding to eigenvalue 1, to appear in the exponent. It was subsequently shown~\cite{Gardi:2013ita} that \emph{all the colour factors appearing in the exponent correspond to connected graphs}. This completes the generalization of the non-Abelian exponentiation theorem to the multi-line case.

\begin{figure}[htb]
\begin{center}
\scalebox{.7}{\includegraphics{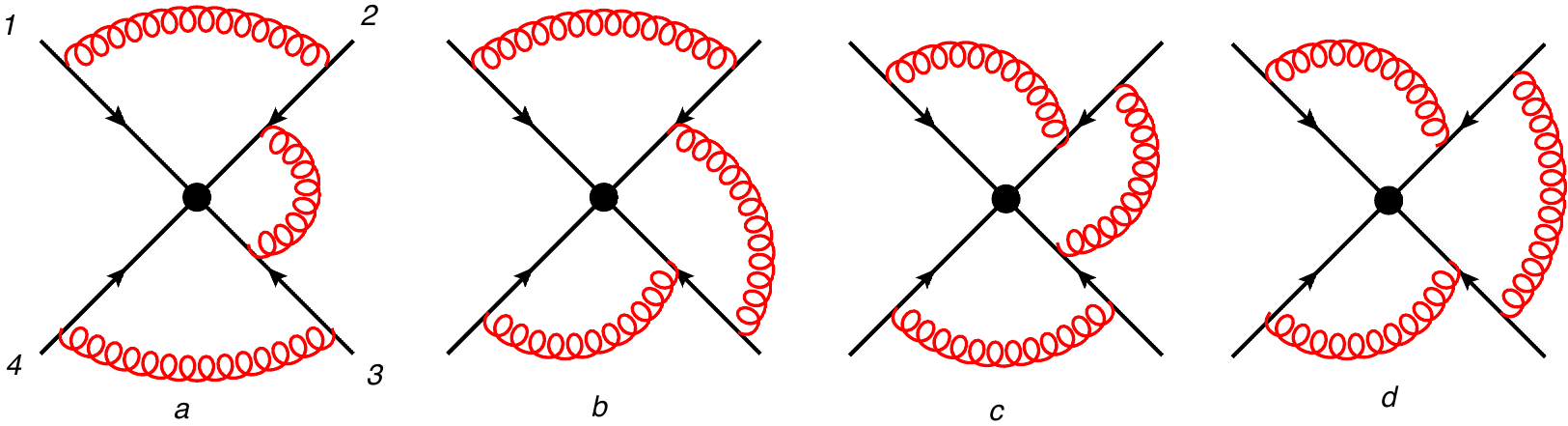}}
\caption{The four diagrams of the multi-gluon-exchange 1-2-2-1 web, all related by permutations of the gluon attachments along lines 3 and 4. The effective colour factor of the web as a whole is connected: $- f^{abe} f^{cde} {\rm T}_1^{a} {\rm T}_2^{b} {\rm T}_3^{c} {\rm T}_4^{d}$. }
\label{1221}
\end{center}
\end{figure}

It is equally important to understand the action of the web mixing matrix on the vector of kinematic integrals ${\cal F}_D$ associated with the diagrams in the web. A step in this direction was taken in \cite{Gardi:2011yz} where it was argued that the linear combinations appearing in the exponent through the action of $R$, for any given web, have the expected structure based on renormalization, as in (\ref{Z}). This is a highly constrained structure, requiring intricate cancellation of the higher-order poles between diagrams. The mixing matrix is responsible for generating this singularity structure, much like the way it generates connected graphs through linear combinations of ordinary colour factors. 

A three-loop example of a web is shown in figure~\ref{1221}: the web 1-2-2-1 consists of the four diagrams whose kinematic integrals are respectively ${\cal F}(a)$ through ${\cal F}(d)$ and their ordinary colour factors are $C(a)$ through $C(d)$, respectively. Eq.~(\ref{Mixing}) in this concrete example is:
{
\begin{align}
\begin{split}
W_{(1,2,2,1)}=&\left(\begin{array}{c}{\cal F}(3a)\\{\cal F}(3b)\\{\cal F}(3c)\\{\cal F}(3d)\end{array}\right)^T\frac{1}{6}\left(\begin{array}{rrrr}1&-1&-1&1\\-2&2&2&-2\\-2&2&2&-2\\1&-1&-1&1\end{array}\right)\left(\begin{array}{c}C(3a)\\C(3b)\\C(3c)\\C(3d)\end{array}\right)\\=&
\underbrace{\frac16 \bigg({\cal F}(a)-2{\cal F}(b)-2{\cal F}(c)+{\cal F}(d)\bigg)}_{\text{subdivergences cancel}}\times \underbrace{\bigg(C(a)-C(b)-C(c)+C(d)\bigg)}_{- f^{abe} f^{cde} {\rm T}_1^{a} {\rm T}_2^{b} {\rm T}_3^{c} {\rm T}_4^{d}}\,.
\end{split}
\end{align}
}
Performing the product we see that this evaluates to one specific linear combination of kinematic integrals multiplying a corresponding combination of colour factors of the original diagrams. Using the colour algebra it is straightforward to see that this colour factor is $- f^{abe} f^{cde} {\rm T}_1^{a} {\rm T}_2^{b} {\rm T}_3^{c} {\rm T}_4^{d}$, which is the same as the connected graph in figure~\ref{4g_connected} (bottom graph). On the kinematic side one observes a cancellation of the cubic pole, while the double pole is associated with commutators of lower order subdiagrams as predicted by the renormalization properties of the Wilson-line vertex (eq.~(\ref{Z})). A detailed analysis of the triple and double pole structure of these diagrams can be found in ref.~\cite{Gardi:2011yz}. The single pole, in turn, contributes to the anomalous dimension -- its calculation~\cite{Gardi:2013saa} will be discussed in the next section.

\section{Multiple gluon exchange webs and their analytic structure\label{sec:mge}}

In the final part of this talk I would like discuss the evaluation of multiple gluon exchange webs such as the 1-2-2-1 web of figure~\ref{1221}. The results for latter along with the 1-1-3-1 web, both of which contribute to the three-loop four-leg soft anomalous dimension,   were published in~\cite{Gardi:2013saa}. The analysis in that paper and in 
the very recent~\cite{Falcioni:2014pka}, goes beyond the computation of specific diagrams: it demonstrates the advantages of organising the calculation in terms of webs, and furthermore defining \emph{subtracted webs}, which are the combinations of webs and commutators of their subdiagrams that enter the anomalous dimension. 
Organised this way, remarkably simply results are obtained, offering some insight into the analytic structure of this entire class of radiative corrections. In the following we briefly summarize these results.

\begin{figure}[htb]
\begin{center}
\scalebox{0.7}{\includegraphics{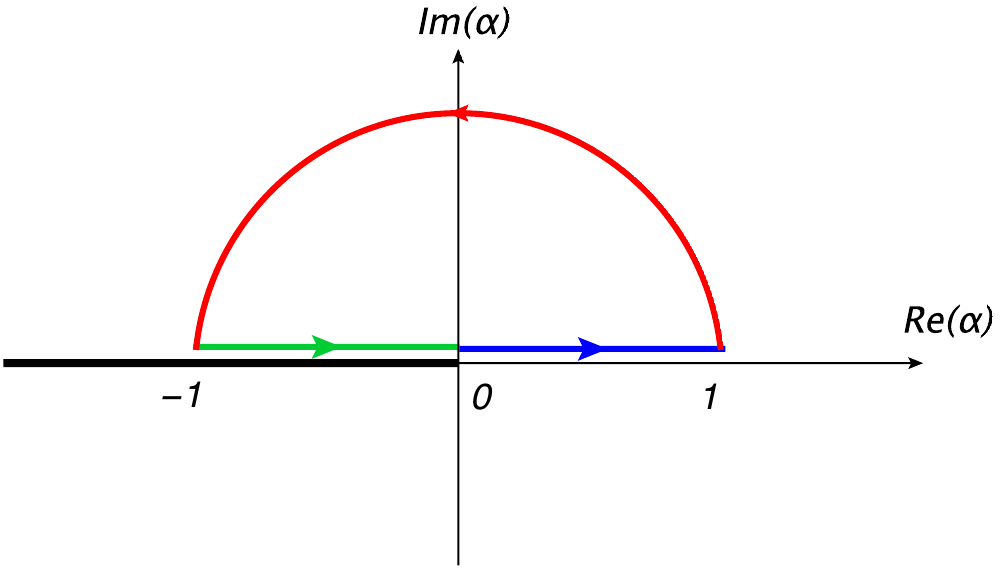}}
\caption{The analytic structure of the one-loop result in the complex $\alpha$ plane -- a logarithmic branch cut along the negative real axis -- shown together with a contour describing the values of $\alpha$ for real values of $\gamma$: the $\alpha\in (0,1)$ region corresponds to space-like kinematics (one incoming and one outgoing parton) where  $\gamma$ varies between $-\infty$ and $-2$;
next, the region of complex $\alpha$ with a positive imaginary part corresponds to the Euclidean region where $-2<\gamma<2$; and finally the region where $\alpha$ is near the branch cut, $\alpha=\alpha_r+{\rm i} \eta$ with $\alpha_r\in (-1,0)$ and $\eta>0$, corresponds to time-like kinematics.  }
\label{alpha_plane}
\end{center}
\end{figure}

Let us begin by performing the last integration in the one-loop calculation of eq.~(\ref{one-loop}) above and examine the result. It is useful to express the result in terms of the kinematic variable $\alpha_{ij}$, corresponding to the exponent of the cusp angle between lines $i$ and $j$. It is related to the previously defined $\gamma_{ij}$ by $\gamma_{ij}=-\alpha_{ij}-1/\alpha_{ij}$. This inherently introduces a symmetry $\alpha_{ij}\to 1/\alpha_{ij}$, so we will only be considering the kinematic region $|\alpha_{ij}|\leq 1$.
It is useful to analyse the physical regions in terms of this kinematic variable. The complex $\alpha_{ij}$ plane is shown in figure~\ref{alpha_plane}. 
The straight-line limit (or Regge limit, where the partons hardly recoil) is $\alpha_{ij}=1$. The region where the invariant $\beta_i\cdot \beta_j$ is spacelike corresponds to $\alpha_{ij}$ between 0 and 1. There is another physical kinematic region where this invariant is timelike and both partons are in the final state or both in the initial state: this is the region where $\alpha_{ij}$ is negative. Finally, the lightlike limit corresponds to $\alpha_{ij}$ near the origin.
The integrand in eq.~(\ref{one-loop}) can be partially-fractioned~\cite{Henn:2013wfa,Gardi:2013saa} as follows:
\begin{align}
\label{p0_part_fracs}
  \hspace*{-8pt}p_0 \left( x, \alpha \right) \, = \, - \, \left( \alpha + \frac{1}{\alpha} \right) \, 
  \frac{1}{x^2 + (1 - x)^2 +\left(\alpha+\frac{1}{\alpha}\right) \, 
  x (1 - x)} \, = \, r (\alpha) \left[ \frac{1}{x - \frac{1}{1 - \alpha}} - 
  \frac{1}{x + \frac{\alpha}{1 - \alpha}} \right] \, ,
\end{align}
where $r(\alpha)=\frac{1+\alpha^2}{1-\alpha^2}$.  Performing the final integration we find $\Gamma_{\cal S}^{(1)}=
\frac{1}{2\pi} \, T_{i}\cdot T_{j} \, F^{(1)}(\alpha_{ij})$ with
\begin{align}
\label{F1}
\begin{split}
F^{(1)}(\alpha_{ij})&= \int_0^1 {\rm d}x \, p_0(x,\alpha_{ij})
= \,\frac{1+\alpha_{ij}^2}{1-\alpha_{ij}^2}
  \int_0^1 {\rm d}x \left(\frac{1}{x - \frac{1}{1 - \alpha_{ij}}} - \frac{1}{x + \frac{\alpha_{ij}}{1 - \alpha_{ij}}} \right)
=\,2\, {\color[rgb]{0.000000,0.000000,0.000000}r(\alpha_{ij})}\,
 \ln\left({\alpha_{ij}}\right)\,.
\end{split}
\end{align}
In a similar manner any diagram involving only gluon exchanges between Wilson lines can be readily brought into a `$d\log$' form, facilitating an algorithmic calculation in terms of polylogarithms~\cite{Henn:2013wfa,Gardi:2013saa}. 
The logarithmic branch cut along the negative real axis we identify in (\ref{F1}) is shown in figure~\ref{alpha_plane}. This corresponds to the physical cut expected for timelike kinematics.
Putting aside the routing of this cut -- or equivalently considering the result at symbol level -- we observe that the final answer exhibits an additional symmetry under $\alpha_{ij}\to -\alpha_{ij}$~\cite{Gardi:2013saa}: this is a crossing symmetry, relating timelike and spacelike kinematics. 
This, as well as the inherent $\alpha_{ij}\to 1/\alpha_{ij}$ symmetry, are consistent with the presence of exactly four singular points in the complex $\alpha_{ij}$ plane, $\{1,-1,0,\infty\}$, which all have a definite physical meaning.  For a quantity such as the cusp anomalous dimension (with massless fields) it is hard to imagine any further physical thresholds developing at higher orders. Indeed it has been observed~\cite{Henn:2012qz,Henn:2013wfa}, that multiple gluon exchange corrections to the cusp anomalous dimension can be expressed in terms of harmonic polylogarithms depending 
on~$\alpha_{ij}^2$. One may then wonder whether this simple analytic structure is preserved at higher orders in the case of multiple Wilson lines.

The two-loop calculation of the 1-2-1 web~\cite{Ferroglia:2009ii,Mitov:2010xw,Chien:2011wz,Gardi:2013saa} already provides a hint that it does. In \cite{Gardi:2013saa} the two-loop three-line soft anomalous dimension coefficient has been expressed as
\begin{align}
\label{Gamma_2_explicit}
\begin{split}
\Gamma_{\cal S}^{(2)}&=-4w^{(2,-1)}_{3g}\underbrace{-4w_{121}^{(2,-1)}-2\left[w^{(1,-1)},w^{(1,0)}\right]}_{\displaystyle -4{\overline{w}_{121}^{(2,-1)}}}\,,
\end{split}
\end{align}
where $w^{(n,k)}$ corresponds to the coefficient of $\alpha_s^n\,\epsilon^k$ of a given web. The two contributions in (\ref{Gamma_2_explicit}) are from the web connecting the three Wilson lines through a three-gluon vertex, $w^{(2,-1)}_{3g}$,  and the 1-2-1 web involving two individual exchanges between the three lines. In eq.~(\ref{Gamma_2_explicit}) the latter has been combined with the commutators of its subdiagrams to form what has been called a subtracted web. This quantity, by construction, is physical, while the non-subtracted web $w_{121}^{(2,-1)}$ on its own is prone to artefacts due to the infrared regulator. Indeed the $\alpha_{ij}\to -\alpha_{ij}$ symmetry is lost for $w_{121}^{(2,-1)}$, but it is recovered for the subtracted web $\overline{w}_{121}^{(2,-1)}$, which evaluates to
\begin{align}
\label{w_sub}
\begin{split}
{\color[rgb]{0.0,0.0,0.0}\overline{w}_{121}^{(2,-1)}}&=- {\rm i}f^{abc} T_i^aT_j^bT_k^c 
\, \frac{1}{(4\pi)^2}\,
\,r(\alpha_{ij})\,r(\alpha_{jk})\Big(\ln(\alpha_{ij})U_1(\alpha_{jk})- \ln(\alpha_{jk})U_1(\alpha_{ij})\Big)\,,
\end{split}
\end{align}
where $U_1$ is a pure dilogarithmic function with the symbol
${\cal S}\left[U_1(\alpha)\right]=-4 \alpha\otimes  \frac{\alpha}{1-\alpha^2}$.

These results motivated a general analysis of the structure of multiple gluon exchange webs~\cite{Gardi:2013saa}. Using the parametrization of the positions of gluon attachments along the Wilson lines introduced in the context of the one-loop calculation in (\ref{one-loop}), now applied for any of the $n$ gluons, namely $\lambda_k=\sigma_k+\tau_k$ and $x_k=\sigma_k/(\sigma_k+\lambda_k)$ we find that the kinematic integrals for any diagram in the web, or indeed for the web as a whole, take the general form:
\begin{align}
\label{genweb} 
  {\cal F}_{W}^{(n)} \left( \gamma_{i j}, \epsilon \right) \, = \,  \kappa^n \, \Gamma( 2 n \epsilon ) 
  \, \int_0^1 \prod_{k = 1}^n \Big[ d x_k  \,\gamma_k \, P_\epsilon \left( x_k, \gamma_k 
  \right) \Big]  \, \phi_{W}^{(n)} \left( x_i; \epsilon \right)  \, , 
\end{align}
where the integration over the distance parameters $\lambda_k$ has been performed, giving rise to the web kernel $\phi_{W}^{(n)} \left( x_i; \epsilon \right)$, while the angular integrations are left undone.  
It has been shown that $\phi_{W}^{(n)} \left( x_i; \epsilon \right)$ is a polylogarithmic function of weight $n-1$. Performing the angular integrations at this point would yield a rather complicated answer, with multiple ultraviolet poles and Goncharov polylogarithms that couple between different 
$\alpha_{ij}$ variables; this result would also invalidate the symbol-level symmetry $\alpha_{ij}\to -\alpha_{ij}$.   However, the favourable way to proceed is to form the subtracted web combination at order $\epsilon^{-1}$, combining the web with the commutators of its subdiagrams, as was done at two loops in eq.~(\ref{Gamma_2_explicit}). This yields the following kinematic dependence for any given colour factor in the subtracted web:
\begin{align}
 \label{subtracted_web_mge_kin}
  F^{(n)}_{W} \big( \alpha_i \big) & =
  \int_0^1 \left[ \, \prod_{k = 1}^n d x_k \, p_0 (x_k, \alpha_k) \right]  \, 
  {\cal G}^{(n)}_{W } \Big(x_i, q(x_i, \alpha_i) \Big) \, \nonumber
 =   \, \left(\prod_{k = 1}^{n} r (\alpha_k) \right)
  G^{(n)}_{W} \big(\alpha_i \big) \, ,
\end{align}
where ${\cal G}^{(n)}_{W }$ is the subtracted web kernel. The key property of the latter, observed in all webs computed so far, is that it is composed exclusively of powers of logarithms of its arguments. All the polylogarithms which appear in individual diagrams, and also in the non-subtracted web kernel $\phi_{W}^{(n)} \left( x_i; \epsilon \right)$ conspire to cancel, thus eliminating any coupling between the different $\alpha_{ij}$ variables through polylogarithms. 

This analysis led to the formulation of the following conjectures concerning multiple-gluon-exchange webs~\cite{Gardi:2013saa}: 
\begin{itemize}
\item{} \emph{Factorization conjecture:} contributions of webs consisting of $n$ gluon exchanges to the soft anomalous dimension (subtracted webs) take the form of a rational factor consisting of a factor of $r(\alpha_{ij})$ for each exchange between lines $i$ and $j$, times a pure polylogarithmic function of weight $2n-1$, which is a sum of products of polylogarithms of individual $\alpha_{ij}$ variables. 
\item{} \emph{Alphabet conjecture:} the symbol alphabet of these polylogarithms is $\{\alpha_{ij},1-\alpha_{ij}^2\}$.
\end{itemize}
It should be stressed that neither the factorization into functions of individual $\alpha_{ij}$ variables, nor the alphabet conjecture have been proven to be general; the violation of one of these elements almost immediately implies violation of the second. Having said that, there is by now significant evidence supporting these conjectures~\cite{Gardi:2013saa,Falcioni:2014pka}, and moreover, an all-weight basis of functions has been constructed~\cite{Falcioni:2014pka}, which is conjectured to be sufficient for expressing any multiple-gluon-exchange web involving any number of Wilson lines at any order.

As an example, we quote the final result for one of the two 4-leg three-loop webs~\cite{Gardi:2013saa}:
\begin{eqnarray}
\begin{split}
{\overline{w}}^{(3,-1)}_{1221} =& -f^{abe}f^{cde}T_1^aT_2^bT_3^cT_4^d\,\,
\left(\frac{1}{4\pi}\right)^3
 \,r(\alpha_{12})r(\alpha_{23})r(\alpha_{34}) \Bigg[-8 U_2(\alpha_{12})\, \ln \alpha_{23}\, \ln \alpha_{34}
\\ & -8 U_2(\alpha_{34}) \,\ln \alpha_{12}\, \ln \alpha_{23}+16\Big(U_2(\alpha_{23})-2\Sigma_2(\alpha_{23})\Big) \,\ln \alpha_{12}\, \ln \alpha_{34}
\\ &-2\ln \alpha_{12} \, U_1(\alpha_{23}) \, U_1(\alpha_{34})
-2\ln \alpha_{34}\, U_1(\alpha_{12}) \, U_1(\alpha_{23})
+4\ln \alpha_{23}\, U_1(\alpha_{12})\, U_1(\alpha_{34})\Bigg]\,.
\end{split}
\end{eqnarray}
This function involves just two simple trilogarithmic functions, whose symbols are
\begin{align}
\begin{split}
{\cal S}\left[U_2(\alpha)\right]&=4\alpha\otimes \frac{\alpha}{1-\alpha^2} \otimes \frac{\alpha}{1-\alpha^2}\,,\\
{\cal S}\left[\Sigma_2(\alpha)\right]&=2\alpha\otimes \alpha\otimes \alpha\,,
\end{split}
\end{align}
consistent with expectations.
One of the most important implications of this structure is that in the light-like limit, this web, and likewise the 1-1-3-1 web, will reduce to products of logarithms of $\alpha_{ij}$, up to terms that are suppressed by powers of $\alpha_{ij}^2$. Thus, these webs -- in contrast to the connected webs of section \ref{sec:connected} -- do not give rise to any finite function of conformally invariant cross ratios. 

\section{Conclusions}

There has been significant progress over the last few years in our understanding, and our ability to compute, long-distance singularities in gauge-theory scattering amplitudes. 
The non-Abelian exponentiation theorem has been generalised to any number of Wilson lines. 
We have seen that this also has practical advantages: we can directly compute the exponent in terms of webs, leading to a major simplification -- in contrast to individual diagrams, subtracted webs share the symmetries and analytic properties of the full anomalous dimension.

We are now very close to completion of the calculation of the soft anomalous dimension for multi-leg scattering in the massless limit. This will finally give a definite answer to a central question posed 5 years ago with regards to corrections to the dipole formula. Much progress has also been made on the calculation of three-loop corrections to the soft anomalous dimension for massive partons. In this process new configuration-space techniques for Wilson-line correlators have been developed, providing insight into the analytic structure of webs.

\bibliographystyle{JHEP}
\bibliography{refs3}

\providecommand{\href}[2]{#2}\begingroup\raggedright\begin{thebibliography}{10}

\bibitem{Becher:2009cu}
T.~Becher and M.~Neubert, {\it {Infrared singularities of scattering amplitudes
  in perturbative QCD}},  {\em Phys. Rev. Lett.} {\bf 102} (2009) 162001,
  [\href{http://xxx.lanl.gov/abs/0901.0722}{{\tt arXiv:0901.0722}}].

\bibitem{Gardi:2009qi}
E.~Gardi and L.~Magnea, {\it {Factorization constraints for soft anomalous
  dimensions in QCD scattering amplitudes}},  {\em JHEP} {\bf 03} (2009) 079,
  [\href{http://xxx.lanl.gov/abs/0901.1091}{{\tt arXiv:0901.1091}}].

\bibitem{Becher:2009qa}
T.~Becher and M.~Neubert, {\it {On the Structure of Infrared Singularities of
  Gauge-Theory Amplitudes}},  {\em JHEP} {\bf 06} (2009) 081,
  [\href{http://xxx.lanl.gov/abs/0903.1126}{{\tt arXiv:0903.1126}}].

\bibitem{Dixon:2009gx}
L.~J. Dixon, {\it {Matter Dependence of the Three-Loop Soft Anomalous Dimension
  Matrix}},  {\em Phys. Rev.} {\bf D79} (2009) 091501,
  [\href{http://xxx.lanl.gov/abs/0901.3414}{{\tt arXiv:0901.3414}}].

\bibitem{Dixon:2009ur}
L.~J. Dixon, E.~Gardi, and L.~Magnea, {\it {On soft singularities at three
  loops and beyond}},  {\em JHEP} {\bf 02} (2010) 081,
  [\href{http://xxx.lanl.gov/abs/0910.3653}{{\tt arXiv:0910.3653}}].

\bibitem{Dixon:2008gr}
L.~J. Dixon, L.~Magnea, and G.~Sterman, {\it {Universal structure of subleading
  infrared poles in gauge theory amplitudes}},  {\em JHEP} {\bf 08} (2008) 022,
  [\href{http://xxx.lanl.gov/abs/0805.3515}{{\tt arXiv:0805.3515}}].

\bibitem{Moch:2005tm}
S.~Moch, J.~Vermaseren, and A.~Vogt, {\it {Three-loop results for quark and
  gluon form-factors}},  {\em Phys.Lett.} {\bf B625} (2005) 245--252,
  [\href{http://xxx.lanl.gov/abs/hep-ph/0508055}{{\tt hep-ph/0508055}}].

\bibitem{Gehrmann:2010ue}
T.~Gehrmann, E.~Glover, T.~Huber, N.~Ikizlerli, and C.~Studerus, {\it
  {Calculation of the quark and gluon form factors to three loops in QCD}},
  {\em JHEP} {\bf 1006} (2010) 094,
  [\href{http://xxx.lanl.gov/abs/1004.3653}{{\tt arXiv:1004.3653}}].

\bibitem{Aybat:2006wq}
S.~M. Aybat, L.~J. Dixon, and G.~F. Sterman, {\it {The two-loop anomalous
  dimension matrix for soft gluon exchange}},  {\em Phys. Rev. Lett.} {\bf 97}
  (2006) 072001, [\href{http://xxx.lanl.gov/abs/hep-ph/0606254}{{\tt
  hep-ph/0606254}}].

\bibitem{Aybat:2006mz}
S.~M. Aybat, L.~J. Dixon, and G.~F. Sterman, {\it {The two-loop soft anomalous
  dimension matrix and resummation at next-to-next-to leading pole}},  {\em
  Phys. Rev.} {\bf D74} (2006) 074004,
  [\href{http://xxx.lanl.gov/abs/hep-ph/0607309}{{\tt hep-ph/0607309}}].

\bibitem{Catani:1998bh}
S.~Catani, {\it {The singular behaviour of {QCD} amplitudes at two-loop
  order}},  {\em Phys. Lett.} {\bf B427} (1998) 161--171,
  [\href{http://xxx.lanl.gov/abs/hep-ph/9802439}{{\tt hep-ph/9802439}}].

\bibitem{Sterman:2002qn}
G.~Sterman and M.~E. Tejeda-Yeomans, {\it Multi-loop amplitudes and
  resummation},  {\em Phys. Lett.} {\bf B552} (2003) 48--56,
  [\href{http://xxx.lanl.gov/abs/hep-ph/0210130}{{\tt hep-ph/0210130}}].

\bibitem{Gardi:2009zv}
E.~Gardi and L.~Magnea, {\it {Infrared singularities in QCD amplitudes}},  {\em
  Nuovo Cim.} {\bf 032C} (2009) 137--157,
  [\href{http://xxx.lanl.gov/abs/0908.3273}{{\tt arXiv:0908.3273}}].

\bibitem{Bret:2011xm}
V.~Del~Duca, C.~Duhr, E.~Gardi, L.~Magnea, and C.~D. White, {\it {An infrared
  approach to Reggeization}},  {\em Phys.Rev.} {\bf D85} (2012) 071104,
  [\href{http://xxx.lanl.gov/abs/1108.5947}{{\tt arXiv:1108.5947}}].

\bibitem{DelDuca:2011ae}
V.~Del~Duca, C.~Duhr, E.~Gardi, L.~Magnea, and C.~D. White, {\it {The Infrared
  structure of gauge theory amplitudes in the high-energy limit}},  {\em JHEP}
  {\bf 1112} (2011) 021, [\href{http://xxx.lanl.gov/abs/1109.3581}{{\tt
  arXiv:1109.3581}}].

\bibitem{Ahrens:2012qz}
V.~Ahrens, M.~Neubert, and L.~Vernazza, {\it {Structure of Infrared
  Singularities of Gauge-Theory Amplitudes at Three and Four Loops}},  {\em
  JHEP} {\bf 1209} (2012) 138, [\href{http://xxx.lanl.gov/abs/1208.4847}{{\tt
  arXiv:1208.4847}}].

\bibitem{Naculich:2013xa}
S.~G. Naculich, H.~Nastase, and H.~J. Schnitzer, {\it {All-loop
  infrared-divergent behavior of most-subleading-color gauge-theory
  amplitudes}},  {\em JHEP} {\bf 1304} (2013) 114,
  [\href{http://xxx.lanl.gov/abs/1301.2234}{{\tt arXiv:1301.2234}}].

\bibitem{Caron-Huot:2013fea}
S.~Caron-Huot, {\it {When does the gluon reggeize?}},
  \href{http://xxx.lanl.gov/abs/1309.6521}{{\tt arXiv:1309.6521}}.

\bibitem{Mitov:2009sv}
A.~Mitov, G.~Sterman, and I.~Sung, {\it {The Massive Soft Anomalous Dimension
  Matrix at Two Loops}},  {\em Phys. Rev.} {\bf D79} (2009) 094015,
  [\href{http://xxx.lanl.gov/abs/0903.3241}{{\tt arXiv:0903.3241}}].

\bibitem{Becher:2009kw}
T.~Becher and M.~Neubert, {\it {Infrared singularities of QCD amplitudes with
  massive partons}},  {\em Phys. Rev.} {\bf D79} (2009) 125004,
  [\href{http://xxx.lanl.gov/abs/0904.1021}{{\tt arXiv:0904.1021}}].

\bibitem{Beneke:2009rj}
M.~Beneke, P.~Falgari, and C.~Schwinn, {\it {Soft radiation in heavy-particle
  pair production: all- order colour structure and two-loop anomalous
  dimension}},  {\em Nucl. Phys.} {\bf B828} (2010) 69--101,
  [\href{http://xxx.lanl.gov/abs/0907.1443}{{\tt arXiv:0907.1443}}].

\bibitem{Czakon:2009zw}
M.~Czakon, A.~Mitov, and G.~F. Sterman, {\it {Threshold Resummation for
  Top-Pair Hadroproduction to Next-to-Next-to-Leading Log}},  {\em Phys. Rev.}
  {\bf D80} (2009) 074017, [\href{http://xxx.lanl.gov/abs/0907.1790}{{\tt
  arXiv:0907.1790}}].

\bibitem{Ferroglia:2009ep}
A.~Ferroglia, M.~Neubert, B.~D. Pecjak, and L.~L. Yang, {\it {Two-loop
  divergences of scattering amplitudes with massive partons}},  {\em Phys. Rev.
  Lett.} {\bf 103} (2009) 201601,
  [\href{http://xxx.lanl.gov/abs/0907.4791}{{\tt arXiv:0907.4791}}].

\bibitem{Ferroglia:2009ii}
A.~Ferroglia, M.~Neubert, B.~D. Pecjak, and L.~L. Yang, {\it {Two-loop
  divergences of massive scattering amplitudes in non-abelian gauge theories}},
   {\em JHEP} {\bf 11} (2009) 062,
  [\href{http://xxx.lanl.gov/abs/0908.3676}{{\tt arXiv:0908.3676}}].

\bibitem{Chiu:2009mg}
J.-y. Chiu, A.~Fuhrer, R.~Kelley, and A.~V. Manohar, {\it {Factorization
  Structure of Gauge Theory Amplitudes and Application to Hard Scattering
  Processes at the LHC}},  {\em Phys. Rev.} {\bf D80} (2009) 094013,
  [\href{http://xxx.lanl.gov/abs/0909.0012}{{\tt arXiv:0909.0012}}].

\bibitem{Mitov:2010xw}
A.~Mitov, G.~F. Sterman, and I.~Sung, {\it {Computation of the Soft Anomalous
  Dimension Matrix in Coordinate Space}},  {\em Phys.Rev.} {\bf D82} (2010)
  034020, [\href{http://xxx.lanl.gov/abs/1005.4646}{{\tt arXiv:1005.4646}}].

\bibitem{Ferroglia:2010mi}
A.~Ferroglia, M.~Neubert, B.~D. Pecjak, and L.~L. Yang, {\it {Infrared
  Singularities and Soft Gluon Resummation with Massive Partons}},
  \href{http://xxx.lanl.gov/abs/1006.4680}{{\tt arXiv:1006.4680}}.

\bibitem{Chien:2011wz}
Y.-T. Chien, M.~D. Schwartz, D.~Simmons-Duffin, and I.~W. Stewart, {\it {Jet
  Physics from Static Charges in AdS}},  {\em Phys.Rev.} {\bf D85} (2012)
  045010, [\href{http://xxx.lanl.gov/abs/1109.6010}{{\tt arXiv:1109.6010}}].

\bibitem{Polyakov:1980ca}
A.~M. Polyakov, {\it {Gauge Fields as Rings of Glue}},  {\em Nucl. Phys.} {\bf
  B164} (1980) 171--188.

\bibitem{Arefeva:1980zd}
I.~Y. Arefeva, {\it {Quantum contour field equstions}},  {\em Phys. Lett.} {\bf
  B93} (1980) 347--353.

\bibitem{Dotsenko:1979wb}
V.~S. Dotsenko and S.~N. Vergeles, {\it {Renormalizability of Phase Factors in
  the Nonabelian Gauge Theory}},  {\em Nucl. Phys.} {\bf B169} (1980) 527.

\bibitem{Brandt:1981kf}
R.~A. Brandt, F.~Neri, and M.-a. Sato, {\it {Renormalization of Loop Functions
  for All Loops}},  {\em Phys. Rev.} {\bf D24} (1981) 879.

\bibitem{Korchemsky:1985xj}
G.~P. Korchemsky and A.~V. Radyushkin, {\it Loop space formalism and
  renormalization group for the infrared asymptotics of {QCD}},  {\em Phys.
  Lett.} {\bf B171} (1986) 459--467.

\bibitem{Correa:2012nk}
D.~Correa, J.~Henn, J.~Maldacena, and A.~Sever, {\it {The cusp anomalous
  dimension at three loops and beyond}},  {\em JHEP} {\bf 1205} (2012) 098,
  [\href{http://xxx.lanl.gov/abs/1203.1019}{{\tt arXiv:1203.1019}}].

\bibitem{Henn:2012qz}
J.~M. Henn and T.~Huber, {\it {Systematics of the cusp anomalous dimension}},
  {\em JHEP} {\bf 1211} (2012) 058,
  [\href{http://xxx.lanl.gov/abs/1207.2161}{{\tt arXiv:1207.2161}}].

\bibitem{Henn:2013wfa}
J.~M. Henn and T.~Huber, {\it {The four-loop cusp anomalous dimension in
  $\mathcal{N} =$ 4 super Yang-Mills and analytic integration techniques for
  Wilson line integrals}},  {\em JHEP} {\bf 1309} (2013) 147,
  [\href{http://xxx.lanl.gov/abs/1304.6418}{{\tt arXiv:1304.6418}}].

\bibitem{Gardi:2013saa}
E.~Gardi, {\it {From Webs to Polylogarithms}},  {\em JHEP} {\bf 1404} (2014)
  044, [\href{http://xxx.lanl.gov/abs/1310.5268}{{\tt arXiv:1310.5268}}].

\bibitem{Falcioni:2014tba}
G.~Falcioni, E.~Gardi, M.~Harley, L.~Magnea, and C.~D. White, {\it {Calculation
  of the 1-1-1-2 web}},  \href{http://xxx.lanl.gov/abs/\rm to appear}{{\tt \rm
  to appear}}.

\bibitem{Almelid:2013tb}
{\O}.~Almelid, C.~Duhr, and E.~Gardi, {\it {Calculation of connected webs in
  configuration space}},  \href{http://xxx.lanl.gov/abs/\rm to appear}{{\tt \rm
  to appear}}.

\bibitem{Korchemsky:1987wg}
G.~P. Korchemsky and A.~V. Radyushkin, {\it {Renormalization of the Wilson
  Loops Beyond the Leading Order}},  {\em Nucl. Phys.} {\bf B283} (1987)
  342--364.

\bibitem{Kidonakis:2009ev}
N.~Kidonakis, {\it {Two-loop soft anomalous dimensions and NNLL resummation for
  heavy quark production}},  {\em Phys. Rev. Lett.} {\bf 102} (2009) 232003,
  [\href{http://xxx.lanl.gov/abs/0903.2561}{{\tt arXiv:0903.2561}}].

\bibitem{Grozin:2014axa}
A.~Grozin, J.~M. Henn, G.~P. Korchemsky, and P.~Marquard, {\it {The $n_{f}$
  terms of the QCD cusp anomalous dimension}},
  \href{http://xxx.lanl.gov/abs/1406.7828}{{\tt arXiv:1406.7828}}.

\bibitem{Goncharov.A.B.:2009tja}
A.~Goncharov, {\it {A simple construction of Grassmannian polylogarithms}},
  \href{http://xxx.lanl.gov/abs/0908.2238}{{\tt arXiv:0908.2238}}.

\bibitem{Goncharov:2010jf}
A.~B. Goncharov, M.~Spradlin, C.~Vergu, and A.~Volovich, {\it {Classical
  Polylogarithms for Amplitudes and Wilson Loops}},  {\em Phys.Rev.Lett.} {\bf
  105} (2010) 151605, [\href{http://xxx.lanl.gov/abs/1006.5703}{{\tt
  arXiv:1006.5703}}].

\bibitem{Duhr:2011zq}
C.~Duhr, H.~Gangl, and J.~R. Rhodes, {\it {From polygons and symbols to
  polylogarithmic functions}},  {\em JHEP} {\bf 1210} (2012) 075,
  [\href{http://xxx.lanl.gov/abs/1110.0458}{{\tt arXiv:1110.0458}}].

\bibitem{Duhr:2012fh}
C.~Duhr, {\it {Hopf algebras, coproducts and symbols: an application to Higgs
  boson amplitudes}},  {\em JHEP} {\bf 1208} (2012) 043,
  [\href{http://xxx.lanl.gov/abs/1203.0454}{{\tt arXiv:1203.0454}}].

\bibitem{Gardi:2013jia}
E.~Gardi, {\it {Progress on soft gluon exponentiation and long-distance
  singularities}},  {\em PoS} {\bf RADCOR2013} (2014) 043,
  [\href{http://xxx.lanl.gov/abs/1401.0139}{{\tt arXiv:1401.0139}}].

\bibitem{Gardi:2010rn}
E.~Gardi, E.~Laenen, G.~Stavenga, and C.~D. White, {\it {Webs in multiparton
  scattering using the replica trick}},  {\em JHEP} {\bf 1011} (2010) 155,
  [\href{http://xxx.lanl.gov/abs/1008.0098}{{\tt arXiv:1008.0098}}].

\bibitem{Mitov:2010rp}
A.~Mitov, G.~Sterman, and I.~Sung, {\it {Diagrammatic Exponentiation for
  Products of Wilson Lines}},  {\em Phys.Rev.} {\bf D82} (2010) 096010,
  [\href{http://xxx.lanl.gov/abs/1008.0099}{{\tt arXiv:1008.0099}}].

\bibitem{Gardi:2011wa}
E.~Gardi and C.~D. White, {\it {General properties of multiparton webs: Proofs
  from combinatorics}},  {\em JHEP} {\bf 1103} (2011) 079,
  [\href{http://xxx.lanl.gov/abs/1102.0756}{{\tt arXiv:1102.0756}}].

\bibitem{Gardi:2011yz}
E.~Gardi, J.~M. Smillie, and C.~D. White, {\it {On the renormalization of
  multiparton webs}},  {\em JHEP} {\bf 1109} (2011) 114,
  [\href{http://xxx.lanl.gov/abs/1108.1357}{{\tt arXiv:1108.1357}}].

\bibitem{Dukes:2013wa}
M.~Dukes, E.~Gardi, E.~Steingrimsson, and C.~D. White, {\it {Web worlds,
  web-colouring matrices, and web-mixing matrices}},
  \href{http://xxx.lanl.gov/abs/1301.6576}{{\tt arXiv:1301.6576}}.

\bibitem{Dukes:2013gea}
M.~Dukes, E.~Gardi, H.~McAslan, D.~J. Scott, and C.~D. White, {\it {Webs and
  Posets}},  \href{http://xxx.lanl.gov/abs/1310.3127}{{\tt arXiv:1310.3127}}.

\bibitem{Gardi:2013ita}
E.~Gardi, J.~M. Smillie, and C.~D. White, {\it {The Non-Abelian Exponentiation
  theorem for multiple Wilson lines}},  {\em JHEP} {\bf 1306} (2013) 088,
  [\href{http://xxx.lanl.gov/abs/1304.7040}{{\tt arXiv:1304.7040}}].

\bibitem{Vladimirov:2014wga}
A.~Vladimirov, {\it {Generating function for web diagrams}},
  \href{http://xxx.lanl.gov/abs/1406.6253}{{\tt arXiv:1406.6253}}.

\bibitem{Yennie:1961ad}
D.~R. Yennie, S.~C. Frautschi, and H.~Suura, {\it {The infrared divergence
  phenomena and high-energy processes}},  {\em Ann. Phys.} {\bf 13} (1961)
  379--452.

\bibitem{Sterman:1981jc}
G.~F. Sterman, {\it Infrared divergences in perturbative {QCD}. (talk)},  {\em
  AIP Conf. Proc.} 22--40.

\bibitem{Gatheral:1983cz}
J.~G.~M. Gatheral, {\it {Exponentiation of eikonal cross-sections in nonabelian
  gauge theories}},  {\em Phys. Lett.} {\bf B133} (1983) 90.

\bibitem{Frenkel:1984pz}
J.~Frenkel and J.~C. Taylor, {\it {Nonabelian eikonal exponentiation}},  {\em
  Nucl. Phys.} {\bf B246} (1984) 231.

\bibitem{Berger:2003zh}
C.~F. Berger, {\it {Soft gluon exponentiation and resummation}},
  \href{http://xxx.lanl.gov/abs/hep-ph/0305076}{{\tt hep-ph/0305076}}. PhD
  Thesis.

\bibitem{Laenen:2008gt}
E.~Laenen, G.~Stavenga, and C.~D. White, {\it {Path integral approach to
  eikonal and next-to-eikonal exponentiation}},  {\em JHEP} {\bf 03} (2009)
  054, [\href{http://xxx.lanl.gov/abs/0811.2067}{{\tt arXiv:0811.2067}}].

\bibitem{Falcioni:2014pka}
G.~Falcioni, E.~Gardi, M.~Harley, L.~Magnea, and C.~D. White, {\it {Multiple
  Gluon Exchange Webs}},  \href{http://xxx.lanl.gov/abs/1407.3477}{{\tt
  arXiv:1407.3477}}.

\end{thebibliography}\endgroup

\end{document}